\documentstyle{article}
\begin{document}
\fontsize{10 pt}{13 pt}
\selectfont
\begin{center}
{\bf ENERGY DISTRIBUTION OF BLACK PLANE SOLUTIONS} \vspace{1 cm}
\fontsize{8pt}{10pt} \selectfont PAUL HALPERN \\
\vspace{2 pt}
{\it Department of Mathematics, Physics and Computer Science,\\
University of the Sciences in Philadelphia, 600 S. 43rd St.,\\
Philadelphia, PA. 19104, USA\\ p.halper@usip.edu}
\end{center}
\begin{quote}
\fontsize{8pt}{10pt} \selectfont We use the Einstein energy-momentum
complex to calculate the energy distribution of static
plane-symmetric solutions of the Einstein-Maxwell equations in 3+1
dimensions with asymptotic anti-de Sitter behavior. This solution is
expressed in terms of three parameters: the mass, electric charge
and cosmological constant. We compare the energy distribution to
that of the Reissner-Nordstr{\"o}m-anti-de Sitter solution, pointing
to qualitative differences between the models. Finally, we examine
these results within the context of the Cooperstock hypothesis.

{\it Keywords:} black hole; black plane; energy-momentum complex.

PACS numbers:  04.20.Cv, 04.40.Nr, 04.70.Bw
\end{quote}
\fontsize{10 pt}{13 pt} \selectfont \vspace{1 cm} \noindent
 {\bf 1. Introduction}

\vspace{6 pt} \noindent In recent years, researchers have extended
the concept of a black hole to include a variety of fields and
geometries.  Various configurations considered include
 two-dimensional black holes \cite{witten}, three-dimensional black holes \cite{banados},
 Kaluza-Klein black holes \cite{wesson}, black strings and black p-branes \cite{gibbons}.
 These models are important in comprehending the relationship between space-time geometry
 and thermodynamic measures such as temperature and entropy.  The goal is to extend the
 Bekenstein-Hawking area law \cite{bekenstein} to include the broadest possible range of structures.
 Exploring these possibilities helps us better understand the causal structure of general relativity,
 as well as its generalizations. Ultimately this could offer insight into the likely nature of a
fully quantized theory of gravitation. Moreover it could point to
avenues of unification--through string theory, M-theory or another
mechanism--of the fundamental natural interactions.

In a significant step, Brill, Louko and Peld{\'a}n \cite{brill} have considered the
thermodynamics of the Reissner-Nordstr{\"o}m-anti-de Sitter solution \cite{carter} in which the two-sphere is
replaced by a two-space of constant negative or vanishing curvature.   They derived Hawking temperatures for these
possibilities, leading to expressions analogous to the first law of black hole thermodynamics.  For each horizon topology,
 they found that the entropy is one-fourth the horizon area.  This extended the Bekenstein-Hawking area law to toroidal
 and other geometries.

Along these lines, one model of particular interest, developed by Rong-Gen Cai and Yuan-Zhong Zhang \cite{cai}, involves
static, plane-symmetric solutions of the Einstein-Maxwell equations with a negative cosmological constant.  These are known as black planes.  Cai and Zhang
found a causal structure similar to the Reissner-Nordstr{\"o}m solution, but with a Hawking temperature that varied with
$M^{1/3}$, where $M$ is the ADM mass density.  They mapped out inner and outer event horizons for these solutions,
identifying how their positions depend on the objects' mass density, charge density, and cosmological constant.
Wu, da Silva, Santos and Wang have further investigated the global structure of such flat topologies \cite{wu}.

In examining static solutions, such as generalizations of black holes, one useful quantity to consider is the
energy distribution.  The localization of energy and momentum is one of the longstanding dilemmas in general relativity.
Defining this physical measure has been a subject of controversy since the foundation of the field.

\fontsize{10 pt}{13 pt} \selectfont \vspace{1 cm} \noindent
 {\bf 2. The Question of the Localization of Energy and Momentum}

\vspace{6 pt} \noindent Einstein himself proposed the first
energy-momentum complex in an attempt to define the local
distribution of energy and momentum \cite{einstein}.  His definition
was followed by many alternative prescriptions, including
formulations by Tolman \cite{tolman} , Papapetrou \cite{papapetrou},
M$\phi$ller \cite{moller}, Landau and Lifshitz \cite{landau},
Weinberg \cite{weinberg} and others.  Being non-tensors under
general coordinate transformations, most of these formulations are
coordinate-dependent, and some have additional restrictions. Some
even yield different results for the same space-time
configuration--vanishing under certain coordinate systems and not in
others.  The sheer variety of such energy-momentum complexes, along
with the formidable nature of the issues concerning them, once led
many researchers to suppose that they weren't well-defined measures.
Some even argued that locally defined energy and momentum had no
meaning within the context of general relativity; only total
energy-momentum could be defined. These concerns stimulated a {\em
quasi-local} approach to energy and momentum, which various
researchers have examined \cite{brown}.

In 1990, however, Virbhadra initiated a re-examination of the subject of energy-momentum complexes \cite{virbhadra1}.
He pointed out that though these complexes are non-tensors, they yield reasonable and consistent results for a given spacetime.
The same year, Bondi demonstrated that energy in general relativity must in principle be localizable \cite{bondi}.

As an important step forward, in 1996 Aguirregabiria, Chamorro and
Virbhadra \cite{acv} found that for any metric of the Kerr-Schild
class, several different definitions of the energy-momentum complex
yield precisely the same results.  (The Kerr-Schild category
includes many well-known solutions, such as Schwarzschild,
Reissner-Nordstr{\"o}m, Kerr-Newman, Vaidya and Bonnor-Vaidya.) Their results suggest a unity and importance to
these various definitions, despite their idiosyncratic dependence on
coordinate choices.

Amongst these varied complexes, Virbhadra has concluded that the
Einstein prescription offers the greatest consistency
\cite{virbhadra2}.  In a detailed study of the question, Xulu has
confirmed this suggestion \cite{xulu}.  Vagenas has similarly
successfully applied the Einstein energy-momentum complex and other
measures to various black hole configurations \cite{vagenas}.

The consistency of various energy-momentum complexes has also been examined in cosmology.
In 1994, Rosen applied the Einstein energy-momentum complex to a closed, isotropic universe,
described by a FRW (Friedmann-Robertson-Walker) metric, and found that the total energy is zero \cite{rosen}.
Another study obtained the same results using the Landau-Lifshitz prescription \cite{johri}.
Yet other work considered the anisotropic Kasner model, and found vanishing total energy using several different
measures \cite{banarjee}.

The question of localization bears significantly upon the study of gravitational waves.  In 1993, Rosen and Virbhadra
applied the Einstein prescription to cylindrical gravitational waves and found the energy and momentum densities to be
finite and reasonable \cite{rosenvirb}.  A further study by Virbhadra found identical results using the prescriptions
of Tolman, as well as Landau and Lifshitz \cite{virbhadra3}.

It is an interesting question whether or not gravitational waves
have energy and momentum content. Cooperstock has hypothesized that
they do not.  For a number of compelling reasons, he has proposed
that, in general, energy and momentum are localized in the regions
of spacetime where the energy-momentum tensor is non-zero
\cite{cooperstock}. This would preclude gravitational waves as
carriers of energy and momentum; instead they would simply convey
information. As Cooperstock points out, such a hypothesis has strong
bearing on the quantum gravity problem, calling into question the
existence of gravitons as elementary particles.  It distinguishes
gravitational disturbances from bosonic fields, such as photons,
that transport energy and momentum in their interactions. He
emphasizes that such a distinction is natural, given that while
other fields interact within a spacetime background, gravitational
waves characterize the dynamics of spacetime itself.

In consideration of these issues, we wish to determine the localized energy distribution of alternative black geometries,
namely black planes, and compare it to that of their spherically symmetric counterparts.  This is part of the overall goal
 of determining which properties are specific to certain geometric configurations, and which are more general.
 By applying the Einstein energy-momentum complex to the Cai-Zhang black plane solution, we hope to ascertain differences
  in the localized energy profile between that and the standard Reissner-Nordstr{\"o}m-anti-de Sitter solution.

\vspace{1 cm} \noindent
 {\bf 3. Plane-symmetric anti-de Sitter Solutions}

\vspace{6 pt} \noindent We start with the black plane metric, with
asymptotically anti-de Sitter behavior.  Following Cai and Zhang's
notation \cite{cai} we can express this as:

\begin{equation}
ds^2 = A(r) dt^2 - {A}^{-1}(r)dr^2 + \frac{\Lambda}{3}r^2(dx^2+dy^2)
\end{equation}
where:
\begin{equation}
A(r) = -\frac{\Lambda}{3}r^2-\frac{m}{r}+\frac{q^2}{r^2}
\end{equation}
with:
\begin{equation}
m=-\frac{12 \pi M}{\Lambda}
\end{equation}
and
\begin{equation}
q=2 \pi Q
\end{equation}

Here, $M$ is the ADM mass, $Q$ is the electric charge and $\Lambda$ is a negative cosmological constant.
Because of the reflection symmetry with respect to the plane $z=0$, we have taken $r= \left| z \right| $.

Now let us calculate the energy distribution of this solution. The
energy-momentum complex in Einstein's prescription is defined as:

\begin{equation}
{{\theta}_i}^k = \frac{1}{16 \pi} {{H_i}^{kl}}_{,l}
\end{equation}

where the superpotentials ${H_i}^{kl}$ are given by:

\begin{equation}
{H_i}^{kl} = \frac{g_{in}}{\sqrt{-g}}{[-g (g^{kn} g^{lm} - g^{ln} g^{km})]}_{,m}
\end{equation}

with the antisymmetric property that:
\begin{equation}
{H_i}^{kl} = -{H_i}^{lk}
\end{equation}

The components of ${H_i}^{kl}$ relevant to the present calculations
are:
\begin{equation}
{H_0}^{03} = {H_0}^{30} = \frac{\Lambda (q^2 - m r - r^4 \frac{\Lambda}{3})}{12 \pi r}
\end{equation}

with the other components of ${H_i}^{kl} = 0$.

The energy-momentum components are found by the volume integral:
\begin{equation}
P_i = \int \int \int {\theta_i}^0 dx^1 dx^2 dx^3
\end{equation}

Through Gauss's theorem we can rewrite this as a surface integral:

\begin{equation}
P_i = \frac{1}{16 \pi} \int \int {H_i}^{0 \alpha}{\mu}_{\alpha} dS
\end{equation}
where ${\mu}_{\alpha}$ is the outward unit vector normal to the surface element $dS$.

If we choose the surface element to be a planar shell with fixed $r$ (and hence two fixed values of $z$),
then $dS = dx dy$ and ${\mu}_{\alpha}$ is the unit radial vector.

Therefore the energy density is:

\begin{equation}
E= P_0 = \frac{{H_0}^{30}}{16 \pi}
\end{equation}

Finally, we substitute expression (8) into this relationship, and
find the energy distribution according to Einstein's prescription
applied to the black plane solution to be:
\begin{equation}
E = M + \frac{\pi \Lambda Q^2} {3 r}  - \frac{{\Lambda}^2 r^3} {36
\pi}
\end{equation}

\vspace{1 cm} \noindent
 {\bf 4. Comparison to the Reissner-Nordstr{\"o}m-anti-de Sitter Solution}

\vspace{6 pt} \noindent It is instructive to compare this
distribution to that of the standard Reissner-Nordstr{\"o}m-anti-de
Sitter solution, with spherical symmetry.  This metric can be
written in the form:
\begin{equation}
ds^2= B(r) dt^2 - B^{-1}(r) dr^2 - r^2 (d{\theta}^2 + {\sin}^2\theta d{\phi}^2)
\end{equation}
where:
\begin{equation}
B(r) = 1 - \frac{2M}{r} + \frac{Q^2}{r^2} - \frac{\Lambda r^2}{3}
\end{equation}
Converting to Kerr-Schild Cartesian form, we define a new time variable T such that:
\begin{equation}
T = t - r + \int {B}^{-1}(r) dr
\end{equation}
The metric now takes the form:
\begin{equation}
ds^2 = dT^2 - dx^2 - dy^2 - dz^2 + (B(r)-1){[ dT + \frac{x dx +y dy +z dz}{r}]}^2
\end{equation}
The energy distribution for the most general nonstatic, spherically
symmetric metric of the Kerr-Schild class was calculated by
Virbhadra \cite{virbhadra2}. Applying that result to the expression
$B(r)$ yields:
\begin{equation}
E = M - \frac{Q^2}{2r} + \frac{\Lambda r^3}{6}
\end{equation}

Let's now assess the differences in the energy distribution
functions for the planar and spherically symmetric geometries.  We
note that for the charged black plane as well as the
Reissner-Nordstr{\"o}m solutions, the energy distribution is
independent of the sign on the charge. Because the curvature depends
on the square of the charge, this is quite expected.

One important distinction between the solutions involves the role of
the cosmological constant in determining the influence of the charge
parameter.  In the black plane case, a positive cosmological
constant would cause energy content to increase with charge. Only
for a negative cosmological constant (a fundamental supposition of
this solution), would energy content decrease with charge. However,
for the Reissner-Nordstr{\"o}m metric, increasing the charge
parameter always decreases the energy content, regardless of the
sign or value of the cosmological constant.

For sufficiently small values of the cosmological constant, the black plane solution would experience a more gradual leveling off of energy density with radial distance than would a Reissner-Nordstr{\"o}m solution of the same mass and charge.  Thus the cosmological constant serves in the former case to modulate the energy distribution.

The black plane solution can be characterized by means of its event horizons.  Cai and Zhang found that the solution,
in general, possesses two horizons. However, they defined an extremal case with only one horizon that occurs if the
following condition is satisfied:
\begin{equation}
Q=\frac{\sqrt{3} M^{2/3} (-\Lambda)^{1/6}}{2 {\pi}^{1/3}}
\end{equation}

For this solution the energy distribution (11) takes the form:
\begin{equation}
M - \frac{{\pi}^{1/3} M^{4/3} {(-\Lambda)}^{4/3}}{4 r} -\frac{r^3 {\Lambda}^2}{36 \pi}
\end{equation}
The horizon for this solution is located at:
\begin{equation}
r = (9 \pi M)^{1/3} (-\Lambda)^{(-2/3)}
\end{equation}
Thus the energy contained inside the horizon is:
\begin{equation}
E = M (\frac{3}{4} -\frac{3^{1/3}{\Lambda}^2}{12})
\end{equation}

For negligibly small $\Lambda$ this is approximately:
\begin{equation}
 E= \frac{3 M}{4}
 \end{equation}

\vspace{1 cm} \noindent
 {\bf 5. Conclusion}

\vspace{6 pt} \noindent Using Einstein's prescription for the
energy-momentum complex, we have investigated the energy
distribution of the black plane model. We found this quantity to be
well-defined and well-behaved. Comparing this case to that of
Reissner-Nordstr{\"o}m-anti-de Sitter solution, we noted that the
charge dependence of the black plane's energy distribution is
regulated by the size of the cosmological constant.

We have also calculated the total energy within the black plane's
event horizon for the extremal case defined by Cai and Zhang.  We
found that if the cosmological constant is small, this energy is
approximately three-quarters of the black plane's mass parameter.

These results appear reasonable, and thus seem to support
Virbhadra's remarks concerning the utility of Einstein's
prescription for a localized definition of energy and momentum. This
picture also tends to support the Cooperstock hypothesis that the
localized energy is zero for regions where the energy-momentum
tensor vanishes. It would be interesting to determine if these
findings are consistent in alternative prescriptions.

\vspace{1 cm} \noindent
 {\bf Acknowledgements}

\vspace{6 pt} \noindent We wish to thank Haverford College for a
visiting professorship during which time much of this work was
completed. We would also like to express our sincere appreciation to
Jonathan Klein for his valuable assistance.  Many thanks to K. S.
Virbhadra for suggesting this topic and for many useful discussions.

\newpage

\vspace{1 cm} \noindent\fontsize{9 pt}{11 pt} \selectfont

\end{document}